\documentclass[prd,twocolumn,showpacs,preprintnumbers,amsmath,amssymb]{revtex4}
\usepackage{graphicx}% Include figure files
\usepackage{epsfig}% Include figure files
\usepackage{rotating}% Include figure files
\usepackage{dcolumn}% Align table columns on decimal point
\usepackage{bm}% bold math

\def\beq     {\begin{equation}}
\def\eeq     {\end{equation}}
\def\bfig    {\begin{figure}}
\def\efig    {\end{figure}}
\def\bcent   {\begin{center}}
\def\ecent   {\end{center}}
\def\btab    {\begin{tabular}}
\def\etab    {\end{tabular}}
\def\barr    {\begin{array}}
\def\earr    {\end{array}}

\def\stop    {\widetilde{t}_{1}}
\def\stst    {\stop {\stop}^*}
\def\rarw    {\rightarrow}
\def\pt      {P_{T}}
\def\met     {\mbox{${E\!\!\!/_T}$}}
\def\mstop   {m_{\stop}}
\def\ttb     {t\bar{t}}
\def\bb      {b\bar{b}}
\def\BR      {BR(\stop\rarw\bar{b}\bar{q})}

\def\etal    {{\em et al.}}

\def\gev     {\: \rm GeV}

\def\fb      {\: \rm fb}
\def\gappeq{\mathrel{\rlap {\raise.5ex\hbox{$>$}}
            {\lower.5ex\hbox{$\sim$}}}}
\def\lappeq{\mathrel{\rlap{\raise.5ex\hbox{$<$}}
            {\lower.5ex\hbox{$\sim$}}}}

\def\slashiii#1{\setbox0=\hbox{$#1$}#1\hskip-\wd0\hbox to\wd0{\hss\sl/\/\hss}}

\begin{document}                                                              
%===================================================================%
\preprint{\large \tt hep-ph/0508009}
\title{Looking for the Top-squark at the Tevatron with four jets}
\author{\sf Debajyoti Choudhury${^1}$, Madhumita Datta${^2}$ and Manas Maity${^2}$}
\affiliation{
$^1${ Department of Physics and Astrophysics, 
University of Delhi,   Delhi 110 007, India}\\
$^2${ Department of Physics, Visva-Bharati,
Santiniketan 731 235, India}}

\begin{abstract}
The scalar partner of the top quark is relatively 
light in many models of supersymmetry breaking. 
We study the production of top squarks (stops) at the Tevatron 
collider and their subsequent decay through baryon-number 
violating couplings such that the final state contains 
no leptons. Performing a detector-level analysis, we demonstrate
that, even in the absence of leptons or missing energy, 
stop masses upto $210 \gev/c^{2}$ can be accessible at the Tevatron.
%PACS numbers: 12.60.Jv, 14.80.Ly, 13.88.+e
\end{abstract}
\maketitle
%===================================================================%
The conservation of either of baryon ($B$) and lepton ($L$) numbers is not
dictated by any fundamental principle and is but an accidental feature of the
perturbative sector of the Standard Model (SM). Indeed, any explanation of the
observed baryon asymmetry of the universe needs at least one of $B$ or $L$ to
be broken to a significant degree. Many extensions of the SM, whether
supersymmetric or not, naturally admit both $B$ and $L$ violation and care
must be taken that both are not violated strongly so as to render the proton
very unstable. As can be easily appreciated, breaking of either $B$ or $L$
would lead to significant alteration in phenomenology, and in particular,
collider signatures for physics beyond the SM. While numerous studies have been
undertaken in the context of $L$ violation, in this article we seek to examine
the experimentally more challenging case of a broken baryon-number.

Low energy supersymmetry ({\sc susy}) is widely considered to be a benchmark
for scenarios going beyond the SM. Since the most general renormalizable
Lagrangian consistent with both gauge symmetries and {\sc susy} contain terms
that break both $B$ and $L$, stability of the proton is normally ensured by
the imposition of an {\em ad hoc} discrete symmetry, namely
$R$-parity~\cite{fayet}.  However, since the same end can be achieved by the
imposition of $L$ alone, we allow, in the superpotential, terms of the
form~\cite{Bviol} \beq W_{R\!\!\!/} = \lambda''_{ijk} \bar{U}^i_R \bar{D}^j_R
\bar{D}^k_R, \label{superpot} \eeq where $\bar{U}^i_R$ and $\bar{D}^i_R$
denote the right-handed up-quark and down-quark superfields respectively.  The
Yukawa couplings $\lambda''_{ijk}$ are antisymmetric under the exchange of the
last two indices. The corresponding Lagrangian can then be written in terms of
the component fields as: 
\beq 
{\cal L}_{R\!\!\!/} = \lambda''_{ijk}
\left(u^c_i d^c_j \tilde{d}^*_k + u^c_i \tilde{d}^*_j d^c_k + \tilde{u}^*_i
  d^c_j d^c_k\right) + {\rm h.c.},
\label{lagrp}
\eeq
thus allowing a squark to decay into a pair of quarks. While 
resonanct production in a hadron collider is possible as 
well~\cite{CDF_dijet}, 
the corresponding rates can be appreciable only if 
two of the superfields belong to the first generation, and then too 
are limited by the size of the couplings $\lambda''$. As can be expected, 
the latter are constrained by various low-energy 
observables~\cite{sher_probir_biswa,bcs_debrupa}, 
though the couplings involving the second and third-generation fields alone 
can be relatively large~\cite{reviews}.  
It is thus advisable to concentrate on the 
(model-independent) strong interactions for squark production and consider 
the effect of $\lambda''$ only in the decays.

In most {\sc susy} models, the large top Yukawa coupling results in the 
the lighter stop, $\tilde t_1$, being light compared to the 
other squarks. Since the realization of the mechanism
of electroweak baryogenesis within the context of the MSSM requires 
light stops, with masses of about or smaller than the top quark 
mass~\cite{EWBG}, there is an added motivation to consider such 
scenarios. 

%------------------------------------------------------------------

At hadron colliders, stop production proceeds overwhelmingly via the strong
interaction and the corresponding cross sections are well known at leading
order \cite{stop_lo}. The next-to-leading order QCD and SUSY-QCD corrections
have been computed \cite{PROSPINOstops} and implemented numerically in {\sc
  prospino}~\cite{PROSPINO,PROSPINOstops}, which we use along with the CTEQ5
parton distribution functions \cite{CTEQ5}.  We further assume that the masses
of the gluino and the other squarks are larger than about 250 GeV so that they
do not alter the NLO cross section significantly\cite{PROSPINOstops}.  This,
furthermore, precludes any significant enhancement of the stop production
cross sections via cascade decays thereby making our estimates {\em
  conservative}.

%------------------------------------------------------------------

The prospects for stop discovery at the Tevatron have been examined both in
the context of $R$-conserving supergravity inspired
scenarios~\cite{stop_tev_sugra} as well as in the context of low-energy {\sc
  susy} breaking\cite{stop_tev_gmsb}.  Search efforts at the 
LEP and the Tevatron,
irrespective of the stop decay mode, have only proved
unsuccessful~\cite{stop_ex}.  The reach, at Run II, depends crucially on the
decay chain (and, hence, the {\sc susy} spectrum) and, for an integrated
luminosity of 2~fb$^{-1}$ typically ranges between 165--190 GeV.  For a stop
light enough such that $\tilde t_1 \to \tilde \chi^+ b$ is kinematically
forbidden, the details of the decay depend very sensitively 
on the mass splitting between
the stop and the lightest neutralino (note that if $R$-parity is broken, the
stop is even allowed to be the lightest {\sc susy} particle).  If, for
example, $m_{\tilde t} < m_W + m_b + m_{\tilde \chi_1^0}$, only two
$R$-conserving decay modes are kinematically accessible, namely ($i$) the
loop-induced flavor-changing two-body decay $\tilde t \to c \tilde \chi^0_1$
and ($ii$) the four-body decay via a virtual $W$ boson, $\tilde t \to W^{+*} b
\tilde \chi_1^0 \to q q b \tilde \chi_1^0$ or $\ell \nu b \tilde \chi_1^0$. It
is easy to see that either of these partial widths are small and may be
superseded by $R$-violating modes even for moderate values of $\lambda''$. For
the rest of this paper, we shall assume that at least one of the modes $\tilde
t_1 \to \bar b + \bar s \, (\bar d)$ has a significant branching fraction. (We
refrain from discussing $\tilde t_1 \to \bar d + \bar s$ for reasons of
experimental sensitivity.)

At this stage, we digress to point out that the stop (or any other
squark) is not the only conjectured strongly-interacting particle that
may decay into a pair of quarks. Even in the simplest
nonsupersymmetric grand unified theories (GUTs), $B$ may be violated
in both the gauge and the scalar sector interactions. The
corresponding elementary particles, namely diquarks, can be either
spin-0 or spin-1 and have baryon and lepton numbers $2/3$ and $0$
respectively~\cite{diquarks}. A generic diquark may transform as ${\bf
3}$ or ${\bf \bar{6}}$ under $SU(3)_c$, as triplet or singlet under
$SU(2)_L$ and can have electric charges $|Q_{\cal D}| = 1/3, 2/3$ or
4/3. Compared to the $\lambda''$s, diquark couplings are typically
less restricted both in terms of symmetry requirements (allowing, for
example, the experimentally easier mode ${\cal D} \to b + b
$~\cite{inprep}) as well as low-energy constraints~\cite{bcs_debrupa}.
As far as scalar diquarks are concerned, a $SU(3)_c$ triplet has the
same production cross section (and phase space distributions) as a
stop of identical mass, while a sextet has a larger one on account of
the color-factor. The cross section for a vector diquark depends on
the exact nature of its gauge interactions and is significantly
larger.  Moreover, a generic diquark tends to decay dominantly into a
pair of quarks. The stop, thus, is the {\em most conservative} choice
from this genre.

At the partonic level, our final state, thus, consists of $ (b \,
q)(\bar b \bar q) $ where $q$ is either a $d$- or a $s$-quark and the
parenthetical pairing is to denote that the combinations arise from the
decay of an (anti-)stop.  The SM backgrounds
were generated with both {\sc madgraph}~\cite{Maltoni:2002qb}
and {\textsc PYTHIA 6.206}\cite{pythia} and tested for consistency.
Using the latter, 
we generate complete events with initial and final state radiation,
multiple interactions, etc., and complete evolution (hadronisation and
decays) of the partons into final state particles.  The latter are
passed through a toy detector simulation (using tools in {\textsc
PYTHIA}) and event reconstruction algorithm mimicking a typical
Tevatron RunII detector. The toy calorimeter has cell sizes of
$\Delta\eta=0.1$ and $\Delta\phi= 15^{\circ}$.  Jet reconstruction has
been done employing the cone algorithm with $\Delta R =
\sqrt{(\Delta\eta)^{2} + (\Delta\phi)^{2}} < 0.7 $ and using calorimeter
clusters with $E_{T} > 1.0 \gev$ as seeds for jet formation.  Only
jets with $|\eta_{jet}| \le 2.4$ and $E_{T} > 15 \gev$ and leptons
with $|\eta_{\ell}| \le 5$ are considered. Tagging of $b$-jets has
been done using decay lengths of $b$-hadrons such that $\sim 60\%$ of
$\ttb$ events have at least one $b$-jet tagged\cite{cdf1}.  Apart from
vertex tagging, soft lepton tag may be used to enhance
$b$-tagging. Since the event features used in this analysis, viz., jet
and lepton $\pt$ and $\eta$, jet multiplicity, $\met$ and $b$-tag are
rather robust and easy to implement, our results would be fairly independent
of the detailed features of a particular detector.

\begin{figure}[!ht]
\epsfig{file=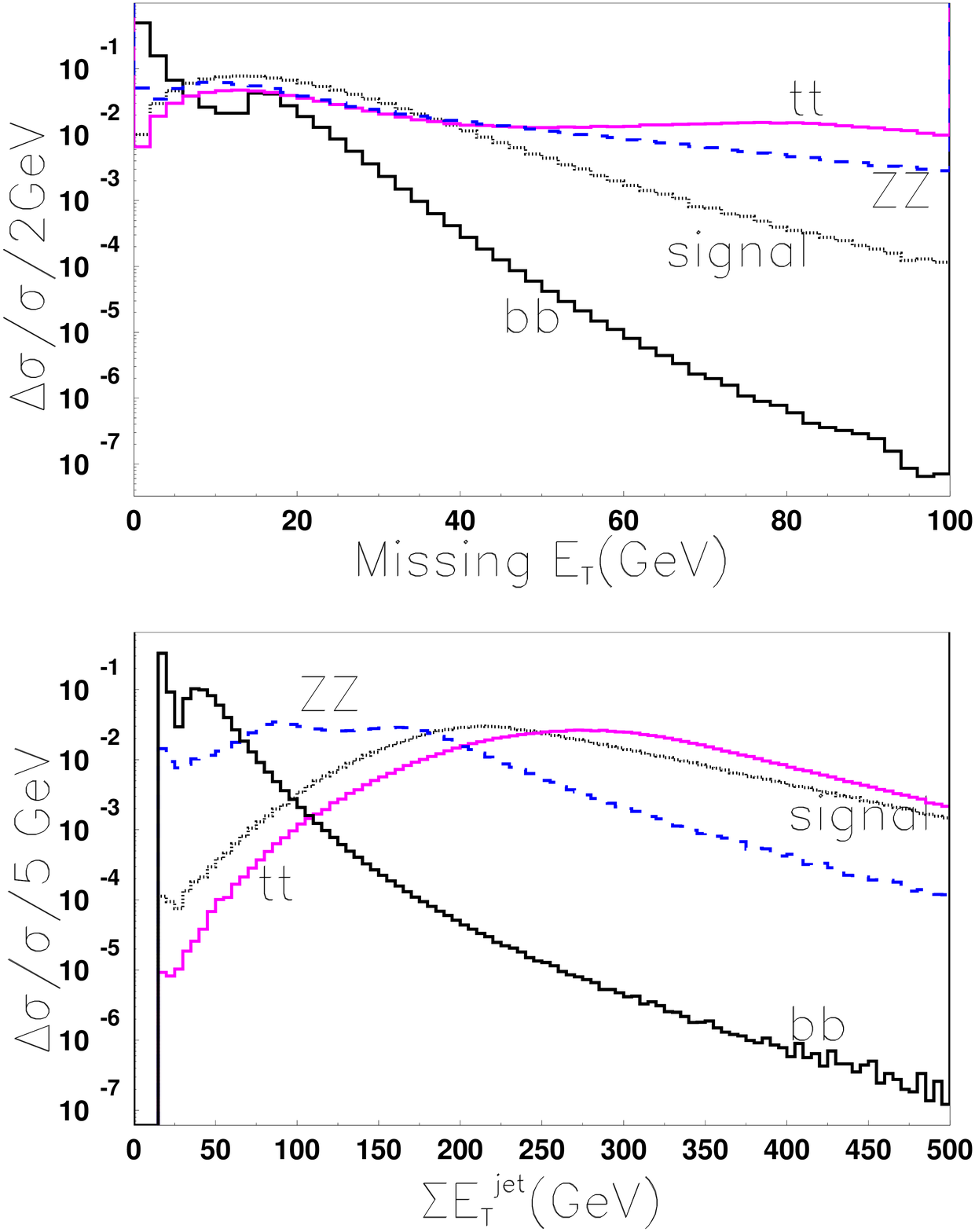,width=8.5cm,height=9.5cm}
\epsfig{file=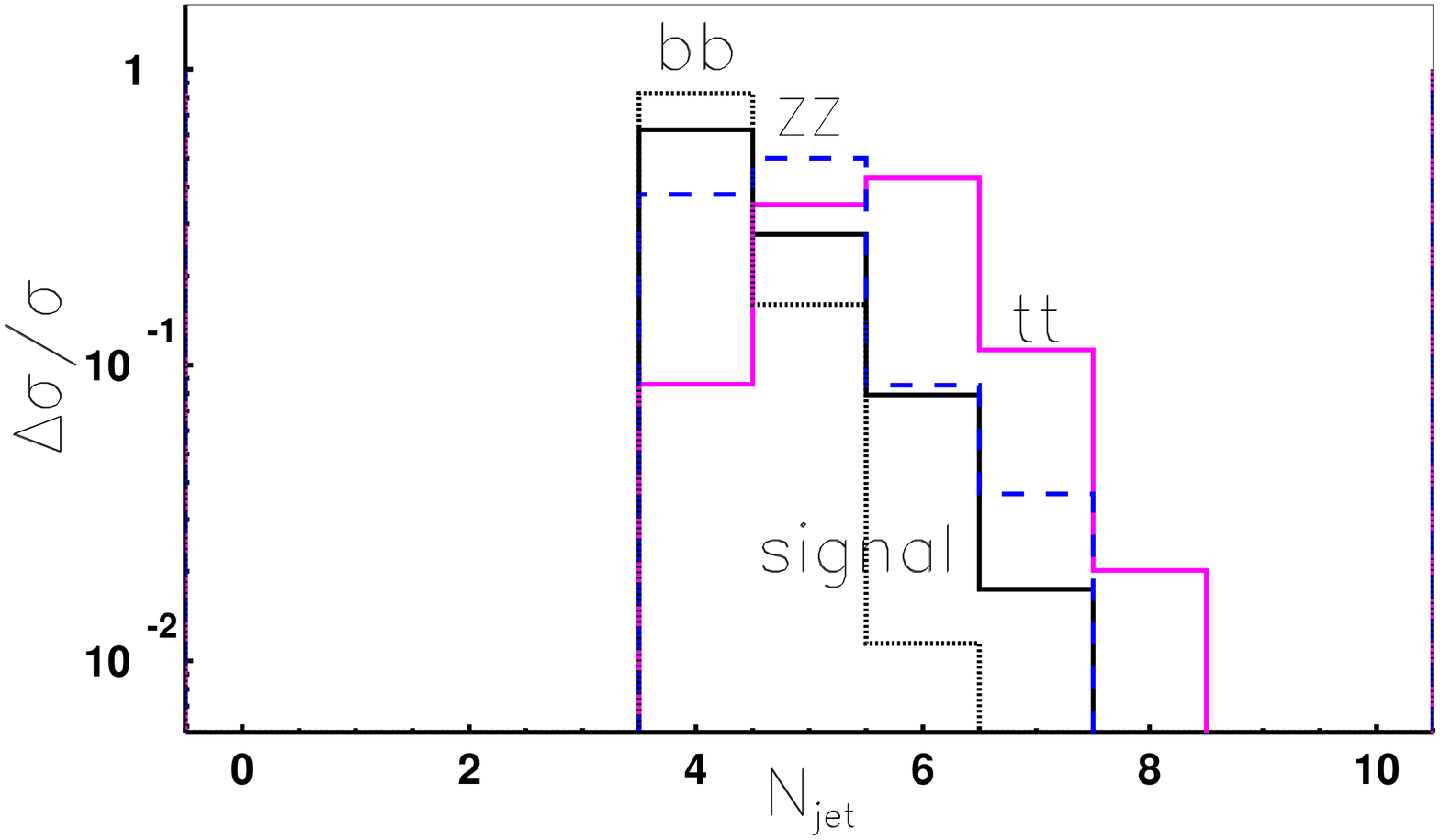,width=8.5cm,height=5.cm}
\caption{\label{fig:fig1}
  Differential distributions (normalized to unity) in different variables for
  both signal ($\mstop = 120 ~GeV/c^{2}$) and various SM backgrounds. The top
  two panels correspond only to the acceptance cuts. For the bottom one, all
  selection critera of eq.(\ref{selection_cuts}) other than that on $N_{\rm
    jet}$ have been imposed as well.}
\end{figure}
The signal events would be characterised 
by four hard jets, two of them being $b$-jets. Since leptons or neutrinos
in such events can occur only as 
decay products of hadrons, these would be soft. This then
inspires our selection criteria
\beq
\barr{lcl}
\met \le 15.0 \gev/c & \quad &
{\rm no \ lepton \ with\ } P_{T}^{\ell} \ge 15 \gev\\[0.5ex]
N_{\rm jet} = 4 & & \sum_{\rm jet} E_{T} > 200 ~\gev \\[0.5ex]
N_{\rm btag} = 2 & & 
M_{bb}, M_{jj} \not \in (70, 100)~{\rm GeV/c^{2}}.
\earr
  \label{selection_cuts}
\eeq

Apart from the $ZZ$ process (which is largely eliminated by the last
requirement above), backgrounds also arise from $\ttb$ events with both tops
decaying hadronically (these typically have more than four jets, see
Fig.~\ref{fig:fig1}) as well as
$\bb$ events accompanied by either or both of multiple interactions and hard
gluon or photon radiation.  Although $\bb$ events have a huge cross-section,
the cut on $\sum_{j} E_{T}^{j}$ is very effective with a rejection factor of
about $10^{4}$ as Fig.~\ref{fig:fig1} amply demonstrates.

This still leaves a large background.  However, in $\stst$ events, of the two
jet pairings viz. $(b_{1}j_{1}, \, b_{2}j_{2})$ and $(b_{1}j_{2}, \,
b_{2}j_{1})$, the one representing the decaying stops should be associated
with only a small difference in the reconstructed invariant masses.  Hence,
our final selection criterion is that \beq |M_{bj}^{1} - M_{bj}^{2}| \le 20
~GeV/c^{2} \eeq for at least one pairing. For the signal events, the
corresponding average of the two masses is expected to show a sharp peak
around $\mstop$ as is evinced by Fig. \ref{fig:recomass}, whereas the other
pairing has a rather flat distribution.

\begin{figure}[!ht]
\epsfig{file=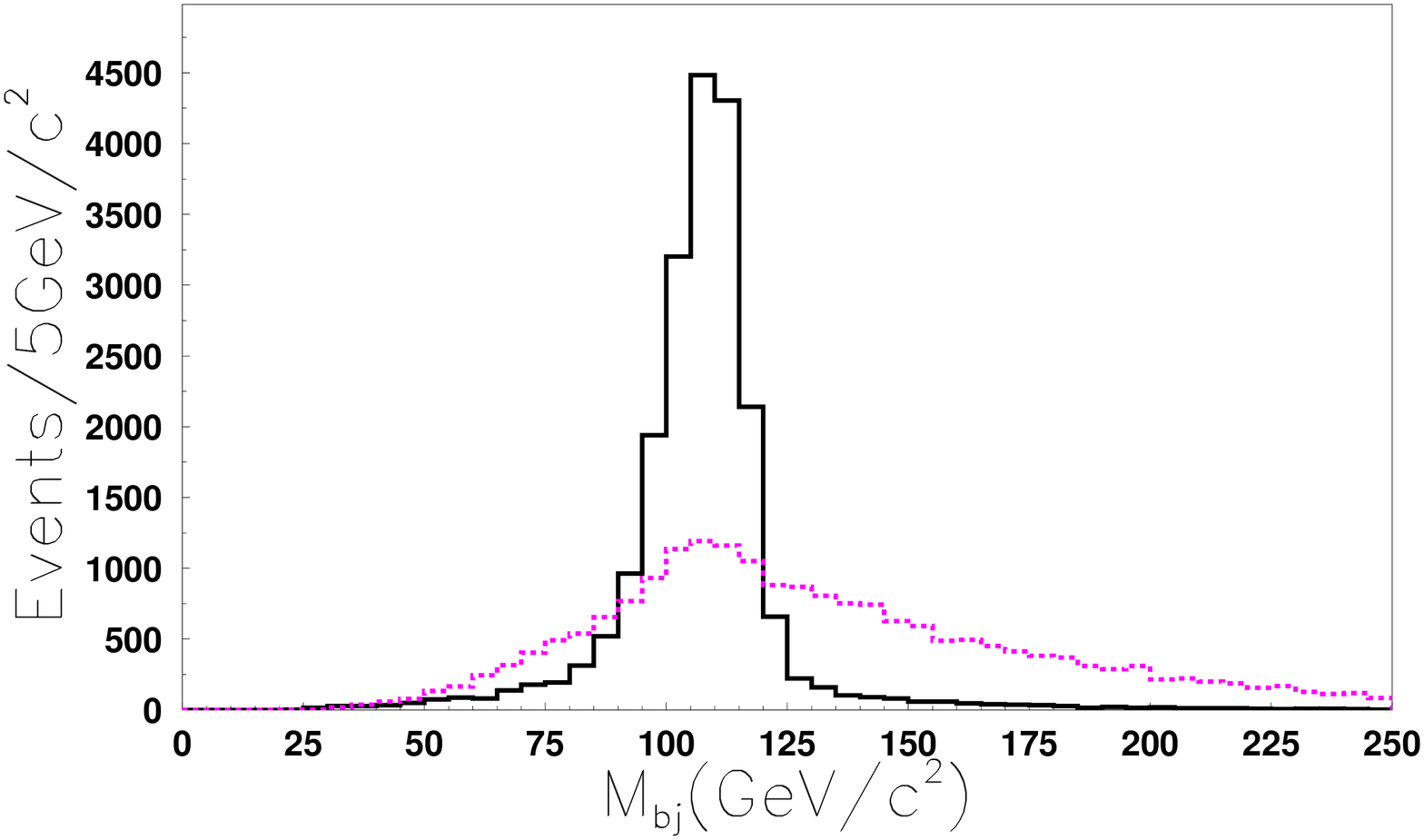,width=8.5cm,height=6.5cm}
\caption{Distribution of the average reconstructed mass for $\stst$ events 
         for $\mstop = 120 ~GeV/c^{2}$ ($10^6$ events generated). 
         The dark (black) line 
         corresponds to the combination with the smaller difference between 
         the two invariant masses; the light (purple) 
         line represents the other  combination.} 
\label{fig:recomass}
\efig

We have simulated $10^{6}$ events for each $\mstop$ and also for 
$\ttb$ and $ZZ$ 
events. Though $\bb$ events have a very small selection efficiency, they 
have a very large cross-section, and constitute the bulk of the 
background events passing the selection cuts. Hence, a very large set 
of $\bb$ events ($\sim 2.5\times 10^{8}$) have been generated  
to get a good estimate of the background distribution. 
As for the signal events, for low $\mstop$, a large fraction of the 
events fail to satisfy the jet selection criteria leading to a small
selection efficiency $\epsilon$ (Fig.\ref{fig:xsec}). As $\mstop$ 
increases, the situation improves rapidly; however beyond $150 ~GeV/c^{2}$,
this effect saturates and is more than offset by the rejection on 
account of hardening of lepton $\pt$ and $\met$. The rapid 
fall in the effective cross-section 
($\sigma \cdot \epsilon$) is, of course, reflective 
of the $p$-wave nature of scalar production.

\begin{figure}[!ht]
\epsfig{file=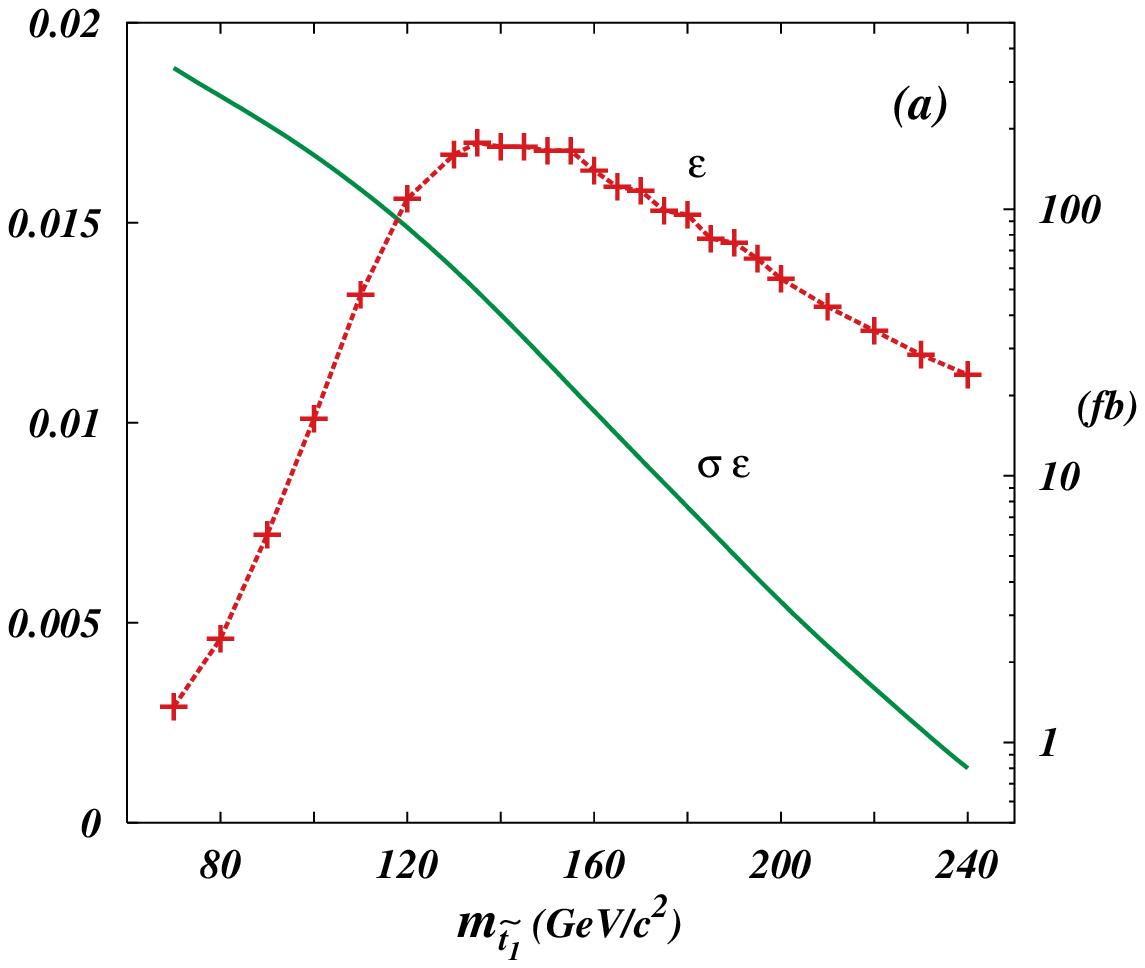,width=8.5cm,height=4.5cm}
\epsfig{file=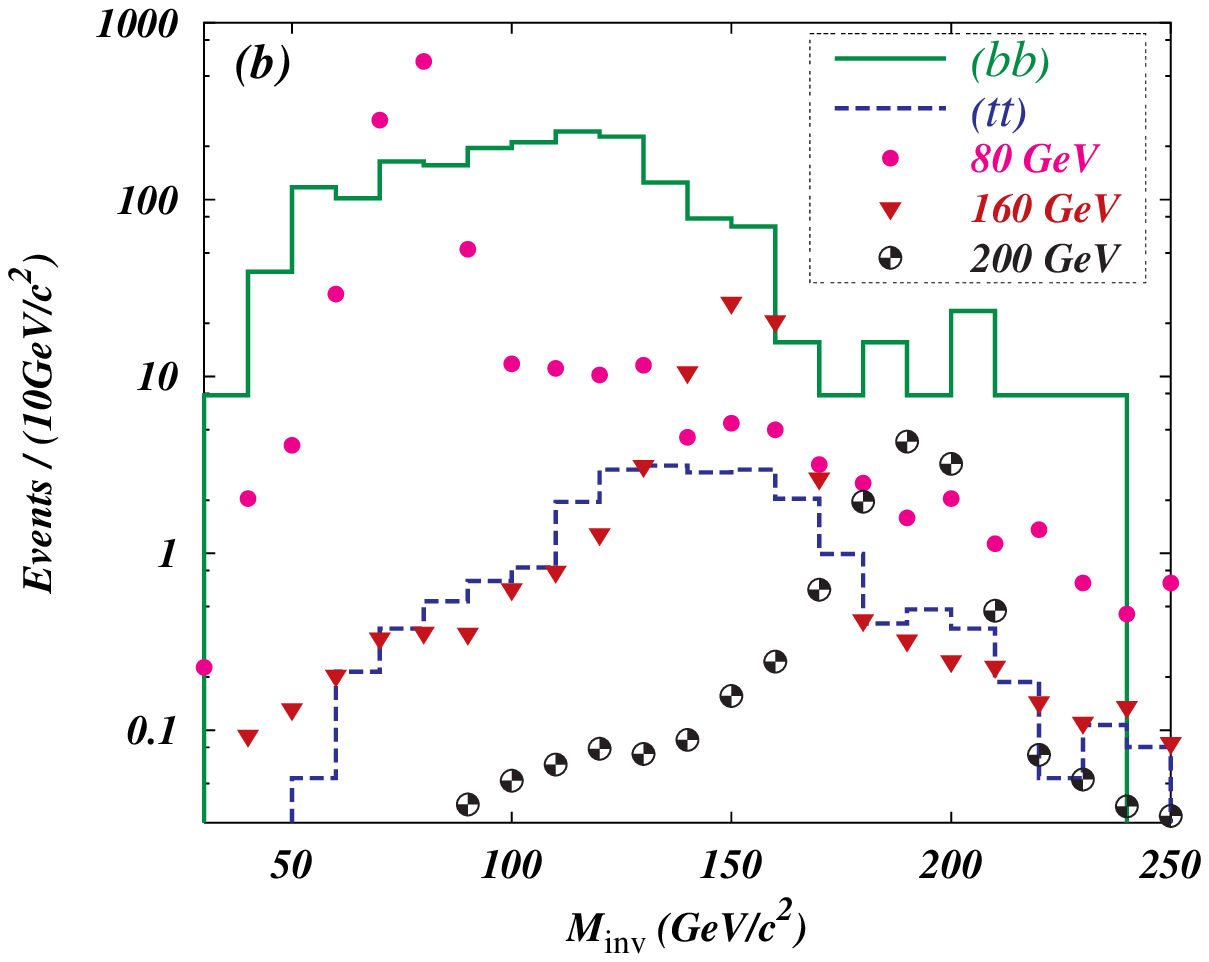,width=8.5cm,height=6.5cm}
\caption{{\em (a)} The detection eficiency for stop-pair production 
and its product with the cross-section as a function of the mass.
{\em (b)} The distribution in the average  of the two masses 
         $M_{bj}$ corresponding to 
         the minimum difference. While the signal profile for 
	 three different stop masses are given by the points, 
	 the solid and dashed lines show the backgrounds from
	 $b \bar b$ (overwhelmingly dominant) and $\ttb$ events.
	 The $ZZ$ rate falls below the scale of the figure.}
\label{fig:xsec}
\end{figure}

Whereas the signal events show a sharp peak in $M_{bj}^{\rm avg}$, the
background is much flatter (Fig. \ref{fig:xsec}).  This allows us to identify
a range in $M_{bj}^{\rm avg}$ where the signal is most significant and
calculate the $\chi^{2}$. Working with a conservative choice of a $50
~GeV/c^{2}$ bin, we use this $\chi^{2}$ to obtain an exclusion plot in the
$\BR-\mstop$ plane (Fig \ref{fig:excl}) that may be reached by the Tevatron
experiments.  With as little as $2 \fb^{-1}$ data, such an analysis would have
a reach upto $185 ~GeV/c^{2}$ (for $\BR = 100\%$), and on the other hand probe
down to $\BR \sim 4\%$ for $\mstop = 70 ~GeV/c^{2}$.  
% This may be extended upto $\sim 2\%$ with ${\cal L}_{int} = 8 \fb^{-1}$.  
Similarly, we may be able to put an upper bound on the $\BR$ for stop
masses upto $200 ~GeV/c^{2}$ with ${\cal L}_{int} = 4 \fb^{-1}$.  A
combined analysis of the data from the two Tevatron RunII experiments
would serve to push the limits even further further.

\begin{figure}[!ht]
\epsfig{file=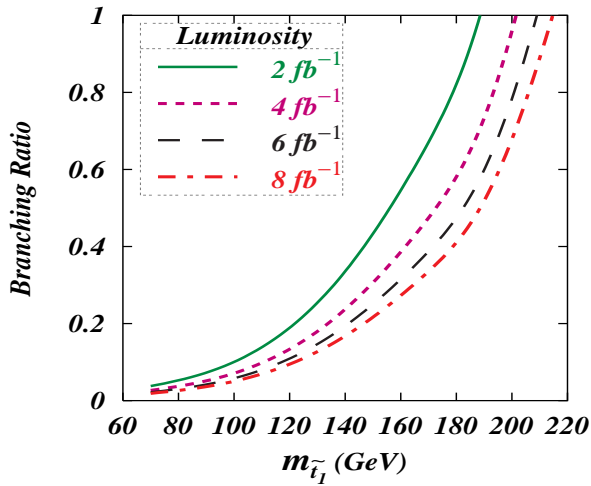,width=14cm,height=8.5cm}
\caption{Exclusion contours at 90\% CL in the $\BR-\mstop$ 
         plane that may be achieved for different values of total integrated luminosity.}
\label{fig:excl}
\end{figure}
In summary, we have outlined above a robust stop-search strategy based on
selection criteria which are easy to implement. Most importantly, 
a final state devoid of leptons or missing energy 
is shown to be very 
promising and  competitive with other modes that have been used
so far. Specific features dependent on detector capabilities may be
used, particularly in a multidimensional analysis, to better
discriminate signal and background and probe added regions in the
parameter space. Furthermore, the sensitivity can be enhanced by considering
$\mstop$-dependent selection criteria rather than the universal cuts that 
we have chosen to impose. In fact, even an analysis of the currently
accumulated data would serve to probe a significant region of the {\sc
susy} parameter space that has not lent itself to an examination so
far.  And as we have already pointed out, the analysis is not limited
to the top-squark or supersymmetry alone but can be readily extended
to diquarks, which, in fact, are generically associated with even
larger cross sections.

DC would like to thank the DST, India for financial assistance 
under the Swarnajayanti Fellowship grant. 
MM would like to thank S. Chakraborty for useful discussions and 
TIFR for the use of computing facilities.

%%%%%%%%%%%%%%%%%%%%%%%%%%%%%%%%%%%%%%%%%%%%%%%%%%%%%%%%%%%%%%%%%%%%%%%%
%                                                                      %
%    Journal macros     (in the Nucl Phys style)                       %
%                                                                      %
%    To change over to Phys. Rev style, replace                        %
%               (#3) #2       by    , #2 (#3)                          %
%                                                                      %
\newcommand{\ib}[3]   {{\em ibid.\/} {\bf #1} (#3) #2}                 %
\newcommand{\app}[3]  {{\em Acta Phys. Polon.   B\/}{\bf #1} (#3) #2}  %
\newcommand{\ajp}[3]  {{\em Am. J. Phys.\/} {\bf #1} (#3) #2}          %
\newcommand{\araa}[3] {{\em Annu. Rev. Astron. Astrophys.\/}           %
          {\bf#1} (#3) #2}                                             %
\newcommand{\apjs}[3] {{\em Astrophys. J. Suppl.\/}                    %
          {\bf  #1} (#3) #2}                                           %
\newcommand{\apjl}[3] {{\em Astrophys. J. Lett.\/} {\bf #1} (#3) #2}   %
\newcommand{\astropp}[3]{Astropart. Phys. {\bf #1} (#3) #2}            %
\newcommand{\eur}[3]  {Eur. Phys. J. {\bf C#1} (#3) #2}                %
\newcommand{\iauc}[4] {{\em IAU Circular\/} #1                         %
       (\ifcase#2\or January \or February \or March  \or April \or May %
                 \or June    \or July     \or August \or September     %
                 \or October \or November \or December                 %
        \fi \ #3, #4)}                                                 %
\newcommand{\ijmp}[3] {Int. J. Mod. Phys. {\bf A#1} (#3) #2}           %
\newcommand{\jetp}[6] {{\em Zh. Eksp. Teor. Fiz.\/} {\bf #1} (#3) #2   %
     [English translation: {\it Sov. Phys.--JETP } {\bf #4} (#6) #5]}  %
\newcommand{\jetpl}[6]{{\em ZhETF Pis'ma\/} {\bf #1} (#3) #2           %
     [English translation: {\it JETP Lett.\/} {\bf #4} (#6) #5]}       %
\newcommand{\jhep}[3] {JHEP {\bf #1} (#3) #2}                          %
\newcommand{\mpla}[3] {Mod. Phys. Lett. {\bf A#1} (#3) #2}             %
\newcommand{\nuovocim}[3]{Nuovo Cim. {\bf #1} (#3) #2}                 %
\newcommand{\np}[3]   {Nucl. Phys. {\bf B#1} (#3) #2}                  %
\newcommand{\npbps}[3]{Nucl. Phys. B (Proc. Suppl.)                    %
           {\bf #1} (#3) #2}                                           %
\newcommand{\philt}[3] {Phil. Trans. Roy. Soc. London A {\bf #1} #2    %
        (#3)}                                                          %
\newcommand{\prev}[3] {Phys. Rev. {\bf #1} (#3) #2}                    %
\newcommand{\plb}[3]  {{Phys. Lett.} {\bf B#1} (#3) #2}                %
\newcommand{\prep}[3] {Phys. Rep. {\bf #1} (#3) #2}                    %
\newcommand{\ptp}[3]  {Prog. Theoret. Phys. (Kyoto) {\bf #1} (#3) #2}  %
\newcommand{\rpp}[3]  {Rep. Prog. Phys. {\bf #1} (#3) #2}              %
\newcommand{\sci}[3]  {Science {\bf #1} (#3) #2}                       %
\newcommand{\zp}[3]   {Z.~Phys. C{\bf#1} (#3) #2}                      %
\newcommand{\uspekhi}[6]{{\em Usp. Fiz. Nauk.\/} {\bf #1} (#3) #2      %
     [English translation: {\it Sov. Phys. Usp.\/} {\bf #4} (#6) #5]}  %
\newcommand{\yadfiz}[4]{Yad. Fiz. {\bf #1} (#3) #2 [English            %
        transl.: Sov. J. Nucl.  Phys. {\bf #1} #3 (#4)]}               %
\newcommand{\hepph}[1] {(electronic archive:    hep--ph/#1)}           %
\newcommand{\hepex}[1] {(electronic archive:    hep--ex/#1)}           %
\newcommand{\astro}[1] {(electronic archive:    astro--ph/#1)}         %
%       \relax                                                         %
%       %%%     End     Journal macro definitions                      %
%                                                                      %
\def\NPB#1#2#3{{Nucl.~Phys.} B {\bf{#1}}, #3 (19#2)}                   %
\def\PLB#1#2#3{{Phys.~Lett.} B {\bf{#1}}, #3 (19#2)}
\def\PRD#1#2#3{{Phys.~Rev.} D {\bf{#1}}, #2 (#3)}
\def\PRL#1#2#3{{Phys.~Rev.~Lett.} {\bf{#1}}, #2 (#3)}
\def\ZPC#1#2#3{{\it Z.~Phys.} {\bf C#1} (19#2) #3}
\def\PTP#1#2#3{{\it Prog.~Theor.~Phys.} {\bf#1}  (19#2) #3}
\def\MPLA#1#2#3{{\it Mod.~Phys.~Lett.} {\bf#1} (19#2) #3}
\def\PR#1#2#3{{\it Phys.~Rep.} {\bf#1} (19#2) #3}
\def\AP#1#2#3{{\it Ann.~Phys.} {\bf#1} (19#2) #3}
\def\RMP#1#2#3{{\it Rev.~Mod.~Phys.} {\bf#1} (19#2) #3}
\def\HPA#1#2#3{{\it Helv.~Phys.~Acta} {\bf#1} (19#2) #3}
\def\JETPL#1#2#3{{\it JETP~Lett.} {\bf#1} (19#2) #3}
%%%%%%%%%%%%%%%%%%%%%%%%%%%%%%%%%%%%%%%%%%%%%%%%%%%%%%%%%%%%%%%%%%%%%%%%

\end{document}